\documentclass[aps,pre,twocolumn,superscriptaddress,showpacs]{revtex4}
\usepackage{graphicx}
\usepackage{amssymb}
\usepackage{epsfig}
\usepackage{amsmath}
\usepackage{times}
\usepackage{bm}

\begin{document}

\title{Controlling synchronous patterns in complex networks}

\author{Weijie Lin}
\affiliation{Department of Physics, Zhejiang University, Hangzhou 310027, China}
\affiliation{School of Physics and Information Technology, Shaanxi Normal University, Xi'an 710062, China}
\author{Huawei Fan}
\affiliation{School of Physics and Information Technology, Shaanxi Normal University, Xi'an 710062, China}
\author{Ying Wang}
\affiliation{School of Physics and Information Technology, Shaanxi Normal University, Xi'an 710062, China}
\author{Heping Ying}
\affiliation{Department of Physics, Zhejiang University, Hangzhou 310027, China}
\author{Xingang Wang}
\email[Email address: ]{wangxg@snnu.edu.cn}
\affiliation{School of Physics and Information Technology, Shaanxi Normal University, Xi'an 710062, China}
\affiliation{Institute of Theoretical \& Computational Physics, Shaanxi Normal University, Xi'an 710062, China}
\begin{abstract}

Although the set of permutation symmetries of a complex network can be very large, few of the symmetries give rise to stable synchronous patterns. Here we present a new framework and develop techniques for controlling synchronization patterns in complex network of coupled chaotic oscillators. Specifically, according to the network permutation symmetry, we design a small-size and weighted network, namely the control network, and use it to control the large-size complex network by means of pinning coupling. We argue mathematically that for \emph{any} of the network symmetries, there always exists a critical pinning strength beyond which the unstable synchronous pattern associated to this symmetry can be stabilized. The feasibility of the control method is verified by numerical simulations of both artificial and real-work networks, and is demonstrated by experiment of coupled chaotic circuits. Our studies pave a way to the control of dynamical patterns in complex networks.

\end{abstract}

\date{\today }
\maketitle

\section{Introduction}

Synchronous behaviors are commonly observed in natural and man-made systems, and are widely recognized as important to the system functionality and operations \cite{SYNBOOK:Kuramoto,SYNBOOK:PRK,SYNBOOK:Strogatz}. In theoretical studies, a popular approach to investigating synchronization is to couple an ensemble of dynamical oscillators, and one of the central tasks is to find the necessary conditions under which the whole system is globally synchronized, or, in some circumstances, the onset of synchronization occurs \cite{SYNREV:Boccaletti,SYNREV:Arenas}. From the viewpoint of synchronization transition, the onset of synchronization and global synchronization stand, respectively, as the starting and ending points, which are mostly concerned for physical and engineering systems \cite{CS:1990,ONSETSYN:WXG,GRIDSYN:Motter}. However, for neuronal and biological systems, experimental evidences have shown that the system dynamics is normally lying somewhere in between \cite{BOOK:Neuron}. Specifically, the oscillators are found to be organized into different clusters, with the motions of the oscillator being highly correlated if they belong to the same cluster, and loosely or not correlated if they belong to different ones \cite{BOOK:Brain}. Stimulated by the experimental observations, in the past decades considerable attentions have been given to the study of cluster (partial, group) synchronization in systems of coupled oscillators \cite{CS:Belykh,CS:Hasler,CS:YZHANG,CS:Pikovsky,CS:BAO,CS:CSZ,CS:OTT2007,CS:Dahms,CS:Nicosia,PATTERNCONTROL:WXG,CS:WXG2014,CS:Pecora2014}.

Recently, with the discoveries of the small-world and scale-free properties in many realistic systems \cite{NET:SW,NET:SFN}, the study of complex network synchronization has received broad interest \cite{NETSYN:CKHU2000,NETSYN:Barahona,NETSYN:Nishikawa2003,NETSYN:Motter2005}. While most of the synchronization phenomena observed in regular networks have been successfully reproduced in complex networks, it remains a challenge to generate cluster synchronization in complex networks, due to the presence of random connections \cite{PATTERN:PARK,PATTERN:WXG2009,PATTERN:WXG2012}. By the bifurcation theory of pattern formation, the generation of cluster synchronization relies strictly on the network symmetry \cite{SYM:Dhys,SYM:Russo,SYM:Golubitsky}, which, at the first glance, is absent in complex networks. In recent years, with the in-depth studies on the relationship between network structure and synchronization, breakthroughs have been made on this topic, and new research interest has been aroused on the study of cluster synchronization in complex networks \cite{CS:BAO,CS:CSZ,CS:OTT2007,CS:Dahms,CS:Nicosia,PATTERNCONTROL:WXG,CS:WXG2014,CS:Pecora2014}. In particular, using the technique of computational group theory \cite{CGT}, in a recent study Pecora \emph{et al.} are able to identify all the permutation symmetries of a complex network and, based on the method of eigenvalue analysis, predict the stable synchronous patterns that can be generated from the random initial conditions \cite{CS:Pecora2014}. This finding changes the conventional picture on cluster synchronization, and, more significantly, points out the ``key" to exploring cluster synchronization in complex networks: the network permutation symmetry \cite{NETSYM}. Although the set of symmetries of a complex network is generally huge, their corresponding synchronous patterns are mostly unstable, making cluster synchronization being hardly observed in complex networks \cite{CS:BAO,CS:CSZ,CS:OTT2007,CS:Dahms,CS:Nicosia,PATTERNCONTROL:WXG,CS:WXG2014,CS:Pecora2014}. Concerning the important implications of cluster synchronization to the security and functioning of many realistic networks, a natural question is: \emph{Is it possible to stabilize these unstable synchronous patterns by some control methods}?

In the present work, we are going to argue mathematically and demonstrate numerically and experimentally that, by a small-size control network designed according to the information of network permutation symmetry, it is indeed possible to control the large-size complex network to \emph{any} synchronous pattern supported by the symmetries. This finding sheds new lights on the synchronous behaviors of complex networks, and equips the control of network dynamics with new technique as well. The rest of the paper is organized as follows. In Sec. II, we shall present our model of networked oscillators, and introduce the new control method. In Sec. III, based on the method of eigenvalue analysis, we shall conduct a detail analysis on the stability of the synchronous patterns, and give the necessary conditions for controlling the patterns. In Sec. IV, we shall apply the control method to different networks, including the numerical studies of artificial and real-world networks, and the experimental study of networked electronic circuits. Discussions and conclusion shall be given in Sec. V.

\section{Model and control method}
Our model of networked chaotic oscillators reads \cite{SYNREV:Boccaletti,SYNREV:Arenas}
\begin{equation}
\dot{\mathbf{x}}_i=\mathbf{F}(\mathbf{x}_i)+\varepsilon\sum\limits^{N}_{j=1}w_{ij}\mathbf{H}(\mathbf{x}_j),
\end{equation}
with $i,j=1,2,\ldots,N$ the oscillator (node) indices and $N$ the network size. $\mathbf{x}_i$ is the state vector associated with the $i$th oscillator, and $\mathbf{F}$ describes the dynamics of the oscillators which, for the sake of simplicity, is set to be identical among the network nodes. $\varepsilon$ is the uniform coupling strength, and $\mathbf{H}(\mathbf{x})$ is the coupling function. The coupling relationship among the oscillators, i.e., the network structure, is captured by the matrix $\mathbf{W}= \{w_{ij}\}$, with $w_{ij}=w_{ji}>0$ the strength that node $j$ is coupled to node $i$. If there is no link between nodes $i$ and $j$, $w_{ij}=0$. The diagonal elements are set as $w_{ii}=-\sum_j w_{ij}$, so as to make $\mathbf{W}$ a Laplacian matrix. Eq. (1), or its equivalent forms, describes the dynamics of a large variety of spatio-temporal systems, which has been employed as one of the standard models in exploring the synchronization behaviors of coupled oscillators \cite{PECORA:Chaos2015}.

Before presenting the new control method, we first describe how to group the network nodes into clusters according to the network symmetries \cite{CS:Pecora2014}. Let $i$ and $j$ be a pair of nodes in the network whose permutation (exchange) does not change the system dynamical equations [Eq. (1)], we call $(i,j)$ a symmetric pair and characterize it by the permutation symmetry $\textsl{g}_{ij}$. Scanning over all the node-pairs in the network, we are able to identify the whole set of permutation symmetries \{$\textsl{g}_{ij}\}$, which forms the symmetry group $\textsl{G}$. Each symmetry $\textsl{g}$ can be further characterized by a permutation matrix $\mathbf{R}_\textsl{g}$, with $r_{ij}=r_{ij}=1$ if $(i,j)$ is a symmetric pair, and $r_{ij}=0$ otherwise. $\mathbf{R}_\textsl{g}$ is commutative with the coupling matrix, $\mathbf{R}_\textsl{g}\mathbf{W}=\mathbf{W}\mathbf{R}_\textsl{g}$, and, after operating on $\mathbf{W}$, it only exchanges the indices of nodes $i$ and $j$. The set of permutation symmetries for a complex network in general is huge \cite{CS:Pecora2014}, but can be obtained from $\mathbf{W}$ by the technique of computational group theory \cite{CGT}. Having obtained the symmetry group $\textsl{G}$, we then can partition the network nodes into clusters according to the permutation orbits, i.e., the subset of nodes permuting among one another by the permutation operations are grouped into the same cluster. In such a way, the network nodes are grouped into a small number of clusters $\{V_l\}$ (see the lower layer in Fig. 1), with $V_l$ the set of nodes belonging to cluster $l$. The clusters provide the topological basis for the formation of synchronous patterns, yet the patterns might not stable, due to either the dynamical or structural instability (more details to be described below).

\begin{figure}[tbp]
\includegraphics[width=0.85\linewidth]{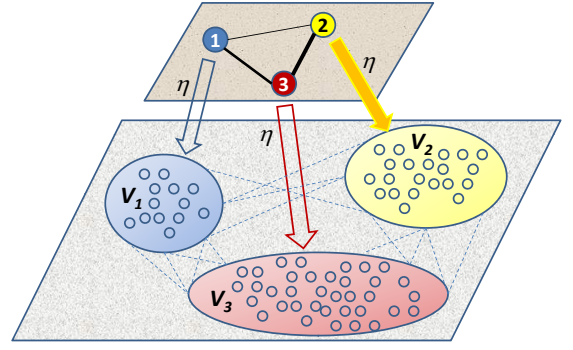}
\caption{(Color online) A schematic plot of the control method. The lower layer represents the network to be controlled, which consists of $M=3$ clusters. Nodes within each cluster are permuting among one another. The upper layer represents the control network, which consists of $3$ controllers. Each controller in the control network is coupled unidirectionally to all nodes in the cluster associated to it.}
\label{Fig1}
\end{figure}

We now present our control method for stabilizing the synchronous patterns. Firstly, a small-size, weighted network is designed according to the cluster information (the upper layer in Fig. 1). The size of the control network is identical to the number of clusters in the original network, and the connections of the control network are weighted as
\begin{equation}
c_{lm}=\sum_{i\in V_{l}}\sum_{j\in V_{m}}w_{ij}/n_{l},
\end{equation}
with $l,m=1,\ldots,M$ the cluster indices and $M$ the size of the control network (also the number of clusters of the original network). $n_l$ represents the size of the $l$th cluster, and $V_{l}$ denotes the set of nodes belonging to cluster $l$. Physically, $c_{lm}$ can be understood as the average coupling strength that a node in cluster $l$ is received from cluster $m$. In general, we have $c_{lm}\neq c_{ml}$, i.e., links in the control network are also directed. The dynamics of the control network is still governed by Eq. (1), except that the coupling matrix is given by Eq. (2) and the network size is changed to $M$. To implement the control, we couple \emph{unidirectionally} each node (controller) in the control network to all nodes in the cluster associated to it. More specifically, the $l$th controller is coupled to all nodes in $V_l$, with $l$ ranging from $1$ to $M$. (A schematic plot of the control method is presented in Fig. 1.)

Unifying the control and original networks into a large-size network, the dynamics of the enlarged network reads
\begin{eqnarray}
\dot{\mathbf{x}}_l&=&\mathbf{F}(\mathbf{x}_l)+\sum^{M}_{m=1}c_{lm}\mathbf{H}(\mathbf{x}_m), \\
\dot{\mathbf{x}}_i&=&\mathbf{F}(\mathbf{x}_i)+\varepsilon\sum\limits^{N}_{j=1}w_{ij}\mathbf{H}(\mathbf{x}_j)
+\eta\varepsilon\sum\limits^{M}_{l=1}\delta_{il}\mathbf{H}(\mathbf{x}_l),
\end{eqnarray}
where Eqs. (3) and (4) describes, respectively, the dynamics of the control and original networks. We set $c_{ll}=-\sum_m c_{lm}$ in Eq. (3), so as to make $\mathbf{C}$ a Laplacian matrix. In Eq. (4), $\eta$ denotes the normalise control (pinning) strength, and $\delta$ is the delta function: $\delta_{il}=1$ if $i\in V_l$, and $\delta_{il}=0$ otherwise. For the system dynamics described by Eqs. (3) and (4), the specific questions we are interested are: Can we stabilize the unstable synchronous patterns by the control network such designed? and, if yes, what is the necessary controlling strength?

\section{Theoretical analysis}

Due to the network symmetry, the topological clusters provide naturally a solution for the synchronous pattern. To be specific, if we set the initial conditions of all nodes inside a cluster to be identical, then during the process of the system evolution, these nodes will be always synchronized whatever the coupling strength. This is because that nodes within the same cluster are surrounded by the same set of neighboring nodes, and thus is perturbed by the same coupling signals. The synchronous pattern such defined, however, is generally unstable. Let $\mathbf{s}_l(t)$ be the synchronous manifold of the $l$th cluster and $\delta \mathbf{x}_i=\mathbf{x}_i-\mathbf{s}_l$ be the infinitesimal perturbations added onto oscillator $i$, then whether the oscillators inside cluster $l$ is synchronizable is basically determined by the following set of variational equations:
\begin{eqnarray}
\delta \dot{\mathbf{x}}_i&=&D\mathbf{F}(\mathbf{s}_l)\delta \mathbf{x}_i
+ \varepsilon \sum_{j\in V_l} w_{ij}D\mathbf{H}(\mathbf{s}_l)\delta \mathbf{x}_j \nonumber\\
&+&\varepsilon \sum_{m\neq l}\sum_{j'\in V_m} w_{ij'}\left[ D\mathbf{H}(\mathbf{s}_m)\delta \mathbf{x}_{j'}-D\mathbf{H}(\mathbf{s}_l)\delta \mathbf{x}_i\right],
\end{eqnarray}
with $i,j=1,\ldots,n_l$ the oscillators belonging to cluster $l$. $\mathbf{s}_m$ denotes the the synchronous manifold of the $m$th cluster, and $\delta \mathbf{x}_{j'}=\mathbf{x}_{j'}-\mathbf{s}_m$, with $j'\in V_m$, represents the perturbations of the $j'$th oscillator from $\mathbf{s}_m$. $D\mathbf{F}(\mathbf{s}_l)$ and $D\mathbf{H}(\mathbf{s}_l)$ are the Jacobin matrices evaluated on $\mathbf{s}_l$. In Eq. (5), the 2nd term on the right-hand-side represents the coupling signals that oscillator $i$ receives from oscillators within the same cluster, and the 3rd term represents the coupling signals that $i$ receives from oscillators in other clusters ($m\neq l$). For the synchronous pattern to be stable, the necessary condition is that $\delta \mathbf{x}_i$ approaches 0 with time for all oscillators in the network. (Please note that this condition is different from that of global synchronization, where all the oscillator trajectories are required to converge to the same manifold, while here oscillators in different clusters are converged to different manifolds.)

Denoting $\Delta\mathbf{X}=[\Delta\mathbf{X}_1,\Delta\mathbf{X}_2,\ldots,\Delta\mathbf{X}_M]^T$ as the network perturbation vector, with $\Delta\mathbf{X}_l=[\delta \mathbf{x}_1,\delta \mathbf{x}_2,\ldots,\delta \mathbf{x}_{n_l}]^T$ the perturbation vector of cluster $l$, then the variational equations of Eq. (5) can be rewritten as
\begin{equation}
\Delta \dot{\mathbf{X}}=\left[ \sum^{M}_{l=1}\mathbf{E}^l\otimes D\mathbf{F}(\mathbf{s}_l)+\varepsilon\mathbf{W}\sum^{M}_{l=1} \mathbf{E}^l\otimes D\mathbf{H}(\mathbf{s}_l)\right]\Delta \mathbf{X}.
\end{equation}
Here, $\mathbf{E}^l$ is an $N$-dimensional diagonal matrix, with $E^{l}_{ii}=1$ if $i\in V_l$, and $E^{l}_{ii}=0$ otherwise. To analyze the stability of the synchronous pattern, the key question is how to decouple the clusters from each other (so that the stability of the clusters can be treated separately). This can be accomplished by transforming the variational equations into the mode space spanned by the eigenvectors of the network permutation matrix, with the details the following. (The mathematical treatment to be described below is modified from Ref. \cite{CS:Pecora2014}, but is more efficient and easy to operate. In Ref. \cite{CS:Pecora2014}, the authors employ the irreducible representations to diagonalize the transverse space of the coupling matrix, which relies on a specially designed code and is computationally costly. In our treatment, we first transform the coupling matrix to the blocked diagonal form, and then diagonalize the transverse blocks separately, which can be done by the conventional routine and is much efficient in simulation.) Firstly, based on the network symmetry, we can construct the network permutation matrix $\mathbf{R}$, with $r_{ij}=r_{ji}=1$ if nodes $i$ and $j$ belong to the same cluster, and $r_{ij}=r_{ji}=0$ otherwise. Secondly, by finding the eigenvectors of $\mathbf{R}$, we can construct the transformation matrix $\mathbf{T}$, such that the transformed matrix $\mathbf{R}'=\mathbf{T}^{-1}\mathbf{R}\mathbf{T}$ is diagonal. Finally, transforming Eq. (6) to the mode space of $\mathbf{T}$, we obtain
\begin{equation}
\Delta \dot{\mathbf{Y}}=\left[ \sum^{M}_{l=1}\mathbf{E}^l\otimes D\mathbf{F}(\mathbf{s}_l)+\varepsilon\mathbf{G}\sum^{M}_{l=1} \mathbf{E}^l\otimes D\mathbf{H}(\mathbf{s}_l)\right]\Delta \mathbf{Y},
\end{equation}
with $\Delta \mathbf{Y}=\mathbf{T}^{-1}\Delta \mathbf{X}$ and $\mathbf{G}=\mathbf{T}^{-1}\mathbf{W}\mathbf{T}$. Particularly, in the mode space the coupling matrix $\mathbf{G}$ has the blocked form
\begin{equation}
\mathbf{G}=\left(
  \begin{array}{ccc}
    \mathbf{B} &      0     \\
         0     &  \mathbf{D} \\
  \end{array}
\right),
\end{equation}
where $\mathbf{B}=\oplus^{M}_{l=1}\mathbf{B}_l$, with $\mathbf{B}_l$ an $(n_l-1)$-dimensional matrix. As $\mathbf{B}_l$ characterizes the motions transverse to the synchronous manifold $\mathbf{s}_l$, we name the associated space the transverse subspace of cluster $l$. The $M$-dimensional matrix $\mathbf{D}$, on the other hand, characterizes the motions parallel to the synchronous manifolds $\{\mathbf{s}_l\}$, we thus name the associated space the synchronous space. Please note that because $\mathbf{W}$ and $\mathbf{G}$ are similar matrices, they have the same set of eigenvalues. In particular, the null eigenvalue, which characterizes the manifold of the global synchronization, belongs to $\mathbf{D}$. The significance of this transformation is that the eigenvalues are now divided into two distinct groups: one for the transverse subspaces (associated to the matrix $\mathbf{B}$) and one for the synchronous subspace (associated to the matrix $\mathbf{D}$). More importantly, the transverse subspaces of the clusters are \emph{decoupled} from each other, so that the synchronization stability of the clusters can be analyzed separately.

\begin{figure*}[tbp]
\begin{center}
\includegraphics[width=0.8\linewidth]{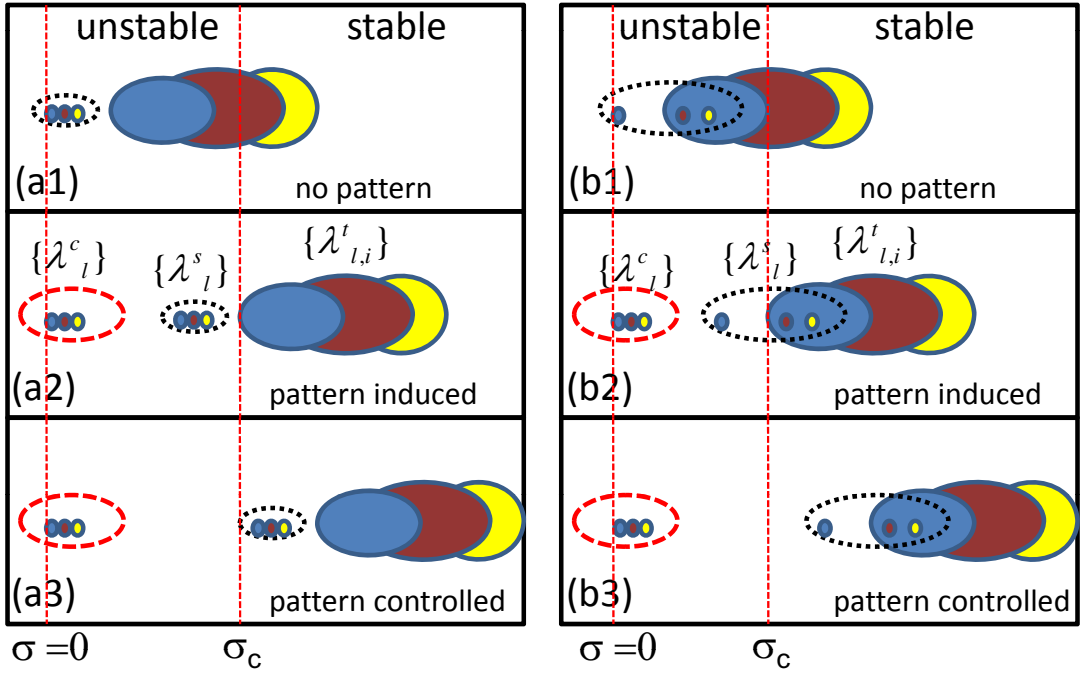}
\caption{(Color online) Schematic plots showing the mechanism of cluster synchronization for the cases of dynamically unstable (a1-a3) and structurally unstable (b1-b3) networks. (a1) Without the control network, the distribution of the eigenmodes in the parameter space of $\sigma$. As some transverse modes are staying in the unstable regime, the synchronous pattern is unstable. (a2) With the control network, synchronous pattern is induced with the controlling strength $\eta>\eta_1$. (a3) $\eta>\eta_2$, the synchronous pattern is controlled to the state of the control network. (b1) Without the control network, the distribution of the eigenmodes. Cluster synchronization is not observable by varying the coupling strength. (b2) Cluster synchronization is induced by the control network when $\eta>\eta_1$. (b3) The synchronous pattern is controlled when $\eta>\eta_2$. Filled colored circles: the transverse modes associated to different clusters. Dotted empty circles (black): the synchronous modes $\{\mathbf{s}_l\}$. Dashed empty circles (red): the controlling modes. Modes are stable in the region $\sigma>\sigma_c$.}
\end{center}
\label{Fig2}
\end{figure*}

Having decoupled the transverse subspaces, the stability of the $l$th cluster is determined by the equation
\begin{equation}
\Delta \dot{\mathbf{Y}}_l=\left[\mathbf{I}_{n'_l} D\mathbf{F}(\mathbf{s}_l)+\varepsilon \mathbf{B}_l D\mathbf{H}(\mathbf{s}_l)\right]\Delta\mathbf{Y}_l,
\end{equation}
with $n'_l=n_l-1$, $\Delta\mathbf{Y}_l=[\delta\mathbf{y}_{l,1},\delta\mathbf{y}_{l,2},\ldots,\delta\mathbf{y}_{l,n'_l}]^T$ the transverse modes of $\mathbf{s}_l$, and $\mathbf{I}_{n'_l}$ the $n'_l$-dimensional identity matrix. To make the $l$th cluster synchronizable, $\delta \mathbf{y}_{l,i}$ should be damping with time for all the $n'_l$ transverse modes. Treating each cluster as an isolated network, this is essentially a problem of global synchronization, which can be analyzed by the method of master stability function (MSF) \cite{MSF-1,MSF-2,MSF-3}. Specifically, projecting $\Delta\mathbf{Y}_l$ into the eigenspace of $\mathbf{B}_l$ (spanned by the eigenvectors of $\mathbf{B}_l$), then Eq. (9) can be transformed into $n'_l$ decoupled equations
\begin{equation}
\delta \dot{\mathbf{y}}_{l,i}=\left[D\mathbf{F}(\mathbf{s}_l)+\varepsilon \lambda^t_{l,i} D\mathbf{H}(\mathbf{s}_l)\right]\delta\mathbf{y}_{l,i},
\end{equation}
with $i\in[1,n'_l]$ and $0>\lambda^t_{l,1}\geq\lambda^t_{l,2}\geq\ldots\geq\lambda^t_{l,n'_l}$ the eigenvalues of $\mathbf{B}_l$. Let $\Lambda_{l,i}$ be the largest Lyapunov exponent calculated from Eq. (10), then whether $\delta \mathbf{y}_{l,i}$ is damping with time is determined by the sign of $\Lambda_{l,i}$: the mode is stable if $\Lambda_{l,i}<0$, and is unstable if $\Lambda_{l,i}>0$. Defining $\sigma\equiv-\varepsilon\lambda^t$, by solving Eq. (10) numerically we can obtain the function $\Lambda=\Lambda(\sigma)$. Previous studies of MSF have shown that for the typical nonlinear oscillators \cite{MSF-1,MSF-2,MSF-3}, $\Lambda$ is negative when $\sigma$ is larger to a critical threshold $\sigma_c$, with $\sigma_c$ a parameter dependent of both the oscillator dynamics and coupling function. [Another typical situation is that $\Lambda<0$ in a bounded region, $(\sigma_1,\sigma_2)$. Our theoretical analysis, as well as the control method, can be generalized to this situation straightforwardly.] As such, to keep the $l$th cluster synchronizable, the necessary condition becomes $\sigma_{l,i}>\sigma_c$ for $i=1,2,\ldots,n'_l$; and, to keep the synchronous pattern stable, this condition should be satisfied for all the $M$ clusters.

To better describe the mechanism of cluster synchronization, we plot Fig. 2, which shows schematically how the transverse modes are stabilized as the coupling (controlling) strength is increased. In Fig. 2, each filled circle represents the transverse modes, $\{\delta \mathbf{y}_l\}$, of a specific cluster, and the dotted empty circle represents the synchronous manifolds, $\{\mathbf{s}_l\}$. Let $0=\lambda^s_{1}>\lambda^s_{2}\geq\ldots\geq\lambda^s_{M}$ be the eigenvalues of $\mathbf{D}$ and denote $\lambda_{min}$ as the largest eigenvalue among all the transverse modes, i.e., $\lambda_{min}=\min\{|\lambda^t_{1,1}|,l=1,\ldots,M\}$, then the scenario of cluster synchronization is dependent of the relationship between $|\lambda^s_{2}|$ and $\lambda_{min}$. If $|\lambda^s_{2}|<\lambda_{min}$ [as the case shown in Fig. 2(a1)], cluster synchronization can be achieved by varying the coupling strength $\varepsilon$. More specifically, given all the transverse modes are staying in the stable region and, in the meantime, at least one of the synchronous modes is staying in the unstable region, cluster synchronization will be emerged. This requirement thus gives the range for generating stable synchronous pattern, $\varepsilon\in(\varepsilon_1,\varepsilon_2)$, with $\varepsilon_1=\sigma_c/\lambda_{min}$ and $\varepsilon_2=\sigma_c/|\lambda^s_{2}|$. On the other hand, if $|\lambda^s_{2}|>\lambda_{min}$ [as the case shown in Fig. 2(b1)], cluster synchronization can not be generated by varying $\varepsilon$. This is because that when the most unstable transverse mode, $\lambda_{min}$, is shifted into the stable region, all the non-trivial synchronous modes will be already in the stable region, resulting in global synchronization instead of cluster synchronization. As $\lambda^s_{2}$ and $\lambda_{min}$ are determined by only the network structure, cluster synchronization can not be generated by varying the coupling strength for the latter, i.e, the synchronous pattern is structurally unstable.

We proceed to analyze the stability of the synchronous pattern when the control network is activated. Regarding the original and control networks as two connected parts of an enlarged network, then the control problem is essentially a problem of network synchronization, except that the synchronous manifolds are predefined by the control network. Similar to the analysis presented above for cluster synchronization, for the control problem we can still decouple the transverse modes by transforming the variational equations into the mode space of the permutation matrix of the enlarged network (which actually is identical to that of the original network, as the controllers are coupled to the oscillators unidirectionally). After the transformation, the coupling matrix has the blocked form
\begin{equation}
\mathbf{G}=\left(
  \begin{array}{ccc}
    \mathbf{B}-\eta\mathbf{I}_{N'} &      0     &   \eta\mathbf{U}\\
         0     &  \mathbf{D}-\eta\mathbf{I}_{M} & \eta\mathbf{I}_{M}\\
         0  &   0   & \mathbf{C}\\
  \end{array}
\right),
\end{equation}
with $\mathbf{B}$ and $\mathbf{D}$ identical to Eq. (8), $\mathbf{C}$ the coupling matrix of the control network, and $\eta$ the controlling strength. $\mathbf{I}_{N'}$ and $\mathbf{I}_M$ are identity matrices of dimensions $N'=N-M$ and $M$, respectively. $\mathbf{U}$ is an $N'\times M$ matrix, with $u_{il}=1$ if $i\in V_{l}$, and $u_{il}=0$ otherwise. From Eq. (11) we see that, with the introduction of the control network, the eigenvalues of $\mathbf{B}$ and $\mathbf{D}$ are increased globally by the amount $\eta$. In particular, $-\lambda_{min}$ and $\lambda^s_1$ are replaced by $-\lambda_{min}-\eta$ and $-\eta$, respectively. If the cluster-synchronization state is dynamically unstable [e.g., Fig. 2(a1)], the necessary condition for generating cluster synchronization becomes $\varepsilon|-\lambda_{min}-\eta|>\sigma_c$ [Fig. 2(a2)], from which we obtain the critical controlling strength for \emph{inducing} cluster synchronization,
\begin{equation}
\eta>\eta_1=(\sigma_c/\varepsilon)-\lambda_{min}.
\end{equation}
Increasing $\eta$ further, the modes associated to $\mathbf{D}$ will be shifted rightward [as depicted in Fig. 2(a3)]. As the controllers are coupled to oscillators in the original network in the one-way fashion, the synchronous manifolds therefore is defined by the control network. That is, the modes associated to $\mathbf{D}$ are switched to transverse modes. Once
\begin{equation}
\eta>\eta_2=\sigma_c/\varepsilon,
\end{equation}
all the transverse modes, $\{\lambda^t_{l,i}\}$ and $\{\lambda^s_l\}$, will be shifted into the stable region, making the cluster-synchronization state being \emph{controlled} to the pattern defined by the control network. If the cluster-synchronization state is structurally unstable [e.g., Fig. 2(b1)], the necessary conditions for inducing and controlling cluster synchronization are still given by Eq. (12) [as depicted in Fig. 2(b2)] and Eq. (13) [as depicted in Fig. 2(b3)], respectively, as the requirement of pattern stability is not changed. We note that the success of generating the structurally unstable pattern lies in the switching of the mode $\lambda^s_1$ from the synchronous to transverse type. That is, the role of $\lambda^s_1$ is replaced by the mode $\lambda^c_1$ of the control network. Here, $0=\lambda^c_1>\lambda^c_2\geq\ldots\geq\lambda^c_M$ are the eigenvalues of $\mathbf{C}$, which are independent of $\eta$.

From the above theoretical analysis, we have the following picture of cluster synchronization. For $\eta<\eta_1$, no cluster is synchronized in the network. Then, as $\eta$ exceeds $\eta_1$, cluster synchronization is emerged in the original network, but is not constrained to the control network. It is worth mentioning that the dynamics of this cluster-synchronization state is identical to that of the control network, i.e., the original network is degenerated to a small-size network which has exactly the same coupling structure as the control network. As a matter of fact, the coupling matrix of the control network [as described by Eq. (2)] is designed based on just such a principle. Finally, as $\eta$ exceeds $\eta_2$, the cluster-synchronization state of the original network is constrained to the sate of the control network, i.e., the synchronous pattern is controlled. Accompanying to this transition, the synchronous manifolds, which are associated to $\mathbf{D}$ when $\eta\in(\eta_1,\eta_2)$, are replaced by the manifolds associated to $\mathbf{C}$.

\section{applications}

To verify the feasibility of the control method, and also to test the theoretical predictions obtained in Sec.III, we next employ the proposed method to control synchronous patterns in different networks, including a small-size network of apparent symmetries, the Nepal power-grid network, and a realistic network of coupled electronic circuits.

\subsection{Small-size network}

We start by applying the control method to a small-size, artificial network. The structure of the network is plotted in Fig. 3(a), which is constructed by a six-node ring network and three nonlocal connections (shortcuts) \cite{CS:WXG2014}. To capture the feature of weighted links widely observed in realistic networks, we set $w_{1,4}=0.8$ for the connection between nodes $1$ and $4$, and $a=1$ for the other connections. In simulations, we adopt the chaotic Lorenz oscillator as the nodal dynamics, which is described by equations $(dx/dt, dy/dt, dz/dt)^{T}=[\alpha(y-x),rx-y-xz,xy-bz]^{T}$. The parameters of the Lorenz oscillator are chosen as $\alpha=10, r=35$, and $b=8/3$, with which the oscillator presents the chaotic motion \cite{Lorenz}. The coupling function is chosen as $\mathbf{H}([x,y,z]^T)=[0,x,0]^T$, i.e., the $x$ variable is coupled to the $y$ variable. Having fixed the nodal dynamics and coupling function, we can obtain the function $\Lambda=\Lambda(\sigma)$ by solving Eq. (10) numerically, which shows that $\Lambda$ is negative when $\sigma>\sigma_c\approx 8.3$ \cite{MSF-3}.

For the simple network presented in Fig. 3(a), the network symmetries can be discerned straightforwardly: the reflection symmetries, $\mathbf{S}_1$ and $\mathbf{S}_2$, and the rotation symmetry (of 180$^\circ$), $\mathbf{S}_3$. As discussed in Sec. III, although each symmetry supports a synchronous pattern, only the stable ones are observable. To figure out the stable synchronous patterns numerically, we investigate the variation of the synchronization relationship among the oscillators as a function of the coupling strength, $\varepsilon$. The results are presented in Fig. 3(b). The synchronization relationship is characterized by the synchronization error $\delta x_i=\langle|x_i-x_2|\rangle$, with $\langle\cdot\cdot\cdot\rangle$ the time average. Fig. 3(b) shows that when $2.6<\varepsilon<4.5$, $\delta x_6=0$ and $\delta x_3=\delta x_5$. That is, two synchronous clusters, $(2,6)$ and $(3,5)$, are formed on the network. Clearly, the synchronized nodes satisfy the symmetry $\mathbf{S}_1$. At $\varepsilon\approx 4.5$, $\delta x_i=0$ for $i=1,\ldots,6$, indicating that the network is globally synchronized. We thus infer from Fig. 3(b) that among the three symmetries, only the synchronous pattern associated to $\mathbf{S}_1$ is observable. In what follows, we are going to demonstrate that, with the help of the control network, the synchronous patterns associated to $\mathbf{S}_2$ and $\mathbf{S}_3$ can also be observed.

\begin{figure*}[tbp]
\begin{center}
\includegraphics[width=0.75\textwidth]{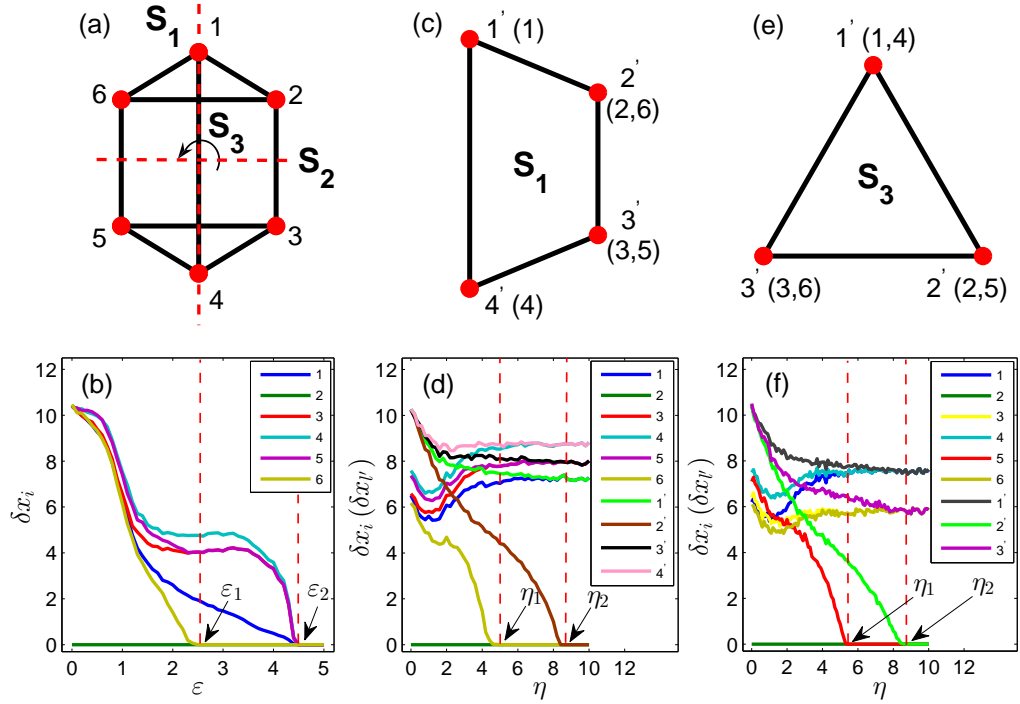}
\caption{(Color online) Controlling synchronous patterns in a six-node network of coupled chaotic Lorenz oscillators. (a) The network structure. The weight of the connection between nodes $1$ and $4$ is 0.8, and is 1 for the other connections. The network possesses two reflection symmetries, $\mathbf{S}_1$ and $\mathbf{S}_2$, and one rotation symmetry, $\mathbf{S}_3$ (180$^\circ$ ration). (b) Without control, the synchronization relationship among the oscillators, characterized by the node synchronization error $\delta x_i=\langle|x_i-x_2|\rangle$, as a function of the coupling strength, $\varepsilon$. Synchronous pattern associated to $\mathbf{S}_1$ is generated in the range $\varepsilon\in(2.6, 4.5)$. (c) The control network designed according to $\mathbf{S}_1$. Controllers $1'$, $2'$, $3'$ and $4'$ control, respectively, the nodes $1$, $(2,6)$, $(3,5)$ and $4$. (d) Fixing $\varepsilon=1.0$, the variation of synchronization relationship among the oscillators (controllers) as a function of the controlling strength, $\eta$. The synchronous pattern associated to $\mathbf{S}_1$ is generated when $\eta>\eta_{1}\approx 5.0$, and is controlled to the control network when $\eta>\eta_{2}\approx8.6$. (e) The control network designed according to $\mathbf{S}_3$. Controllers $1'$, $2'$ and $3'$ control, respectively, the symmetric pairs $(1,4)$, $(2,5)$ and $(3,6)$. (f) Fixing $\varepsilon=1.0$, the variation of synchronization relationship among the oscillators (controllers) as a function of $\eta$. The synchronous pattern associated to $\mathbf{S}_3$ is generated when $\eta>\eta_{1}\approx 5.7$, and is controlled to the control network when $\eta>\eta_{2}\approx8.6$. In (d) and (f), $\delta x_{l'}=\langle|x_{l'}-x_2|\rangle$, with $l'$ the controller index.} \label{Fig3}
\end{center}
\end{figure*}

To study, we set $\varepsilon=1.0<\varepsilon_1$, with which no synchronous cluster is formed. Our first attempt is to generate the synchronous pattern associated $\mathbf{S}_1$. (Although this state is structurally stable, it is dynamically unstable due to the weak coupling.) According to the symmetry $\mathbf{S}_1$, we construct the control network shown in Fig. 3(c). According to Eq. (2), we have the coupling matrix of the control network
\begin{equation}
\mathbf{C}=\left(
  \begin{array}{cccc}
      -2.8&    2&   0&   0.8\\
         1&   -2&   1&   0 \\
         0&    1&  -2&   1 \\
       0.8&    0&   2&  -2.8\\
  \end{array}
\right).
\end{equation}
In implementing the control, controllers $1'$, $2'$, $3'$ and $4'$ are coupled unidirectionally to nodes $1$, $(2,6)$, $(3,5)$ and $6$ in the original network, respectively. By solving the Eqs. (3) and (4) numerically, we plot in Fig. 3(d) the variation of the synchronization relationship among the oscillators as a function of the controlling strength, $\eta$. It is seen that at $\eta_1\approx 5.0$, the synchronous pattern of $\mathbf{S}_1$ is successfully induced ($\delta x_2=\delta x_6=0$ and $\delta x_3=\delta x_5$); and, at $\eta_2\approx 8.6$, the synchronous pattern is controlled ($\delta x_2=\delta x_6=\delta x_{2'}=0$, $\delta x_3=\delta x_5=\delta x_{3'}$, $\delta x_1=\delta x_{1'}$, and $\delta x_4=\delta x_{4'}$), i.e., nodes $2$ and $6$ (nodes $3$ and $5$) are synchronized to controller $2'$ (controller $3'$).

The critical controlling strengths, $\eta_1$ and $\eta_2$, can be analyzed by the theory proposed in Sec. III, as follows. Corresponding to the network symmetry $\mathbf{S}_1$, we have the permutation matrix
\begin{equation}
\mathbf{R}=\left(
  \begin{array}{cccccc}
    1 & 0 & 0 & 0 & 0 & 0\\
    0 & 0 & 0 & 0 & 0 & 1\\
    0 & 0 & 0 & 0 & 1 & 0\\
    0 & 0 & 0 & 1 & 0 & 0\\
    0 & 0 & 1 & 0 & 0 & 0\\
    0 & 1 & 0 & 0 & 0 & 0\\
  \end{array}
\right),
\end{equation}
from which we can obtain transformation matrix (constructed by the eigenvectors of $\mathbf{R}$), which reads
\begin{equation}
\mathbf{T}=\left(
  \begin{array}{cccccc}
      0  &   0  &  1  &   0 & 0  &  0\\
   0  &  0.7071   &   0  & 0  &  0.7071  &0\\
   -0.7071  & 0  & 0  & 0  & 0  &  0.7071\\
         0  & 0  &  0   & 1 & 0   & 0\\
    0.7071  &   0   &   0  &    0 & 0 & 0.7071\\
         0  & -0.7071 &  0  &  0  &  0.7071 &   0\\
  \end{array}
\right).
\end{equation}
The coupling matrix of the original network is
\begin{equation}
\mathbf{W}=\left(
  \begin{array}{cccccc}
    -2.8 & 1 & 0 & 0.8 & 0 & 1 \\
    1 & -3 & -1 & 0 & 0 & 1  \\
    0&1	&-3&1&	1	&0\\
    0.8 & 0	& 1	&-2.8 &	1 & 0\\
    0	&0	&1&	1	&-3&	1\\
    1	&1	&0	&0&	1&	-3\\
  \end{array}
\right)
\end{equation}
which, after the transformation operation $\mathbf{G}=\mathbf{T}^{-1}\mathbf{W}\mathbf{T}$, has the blocked form shown in Eq. (8), with
\begin{equation}
\mathbf{B}=\left(
  \begin{array}{cc}
      -4 & -1 \\
      -1 & -4 \\
  \end{array}
\right)
\end{equation}
and
\begin{equation}
\mathbf{D}=\left(
  \begin{array}{cccc}
      -2.8&   0.8&   1.4142  &    0\\
    0.8&   -2.8&        0 &   1.4142\\
    1.4142    &  0  & -2&   1\\
         0 &   1.4142 &   1&   -2\\
  \end{array}
\right)
\end{equation}
For the matrix $\mathbf{B}$ (which characterizes the transverse spaces of the synchronous pattern), the eigenvalues are $(\lambda^{t}_1,\lambda^{t}_2)=(-3, -5)$; for the matrix $\mathbf{D}$ (which characterizes the synchronous spaces of the synchronous pattern), the eigenvalues are $(\lambda^{s}_1,\lambda^{s}_2,\lambda^{s}_3,\lambda^{s}_4)=(0, -1.85, -3, -4.75)$. From Eqs. (12) and (13), we therefore have $\eta_1=(\sigma_c/\varepsilon)-\lambda_{min}=8.3-|\lambda^t_1|=5.3$ and $\eta_2=\sigma_c/\varepsilon=8.3$, which are in good agreement with the numerical results shown in Fig. 3(d).

Our second attempt is to generate the synchronous pattern associated to $\mathbf{S}_3$, which, according to the our definitions in Sec. III, is classified as structurally unstable (i.e., $\lambda_{min}|<\lambda^s_2|$). The control network is presented in Fig. 3(e), in which controllers $1'$, $2'$ and $3'$ are used to control the symmetric pairs $(1,4)$, $(2,5)$ and $(3,6)$, respectively. The coupling matrix of the control network reads
\begin{equation}
\mathbf{C}=\left(
  \begin{array}{ccc}
      -2&   1&    1\\
       1&  -3&    2\\
       1&   2&   -3\\
  \end{array}
\right).
\end{equation}
Fig. 3(f) shows the variation of the synchronization relationship of the oscillators as a function of the controlling strength, $\eta$. It is shown that when $\eta > \eta_1\approx 5.7$, we have $\delta x_1=\delta _4$, $\delta x_2=\delta x_6$ and $\delta x_3=\delta x_5$. That is, the synchronous pattern defined by $\mathbf{S}_3$ is generated. Increasing $\eta$ further to $\eta_2\approx 8.6$, we have $\delta x_{1'}=\delta x_1=\delta _4$, $\delta x_{2'}=\delta x_2=\delta x_6$ and $\delta x_{3'}=\delta x_3=\delta x_5$, i.e., the synchronous pattern is controlled.

Still, the critical controlling strengths, $\eta_1$ and $\eta_2$, can be analyzed by the theory presented in Sec. III. For the network symmetry $\mathbf{S}_3$, we have the permutation matrix
\begin{equation}
\mathbf{R}=\left(
  \begin{array}{cccccc}
    0 & 0 & 0 & 1 & 0 & 0\\
    0 & 0 & 0 & 0 & 1 & 0\\
    0 & 0 & 0 & 0 & 0 & 1\\
    1 & 0 & 0 & 0 & 0 & 0\\
    0 & 1 & 0 & 0 & 0 & 0\\
    0 & 0 & 1 & 0 & 0 & 0\\
  \end{array}
\right).
\end{equation}
Transforming the coupling matrix $\mathbf{W}$ [Eq. (17)] into the mode space spanned by the eigenvectors of $\mathbf{R}$, we have the blocked matrix $\mathbf{G}$ [of the form shown in Eq. (8)], with
\begin{equation}
\mathbf{B}=\left(
  \begin{array}{ccc}
      -3 &  1   &  0\\
       1 & -3.6 & -1\\
       0 & -1   & -3\\
  \end{array}
\right),
\end{equation}
and
\begin{equation}
\mathbf{D}=\left(
  \begin{array}{cccc}
      -3 &  2 &  1\\
       2 & -3 &  1\\
       1 &  1 & -2\\
  \end{array}
\right).
\end{equation}
For the matrix $\mathbf{B}$, we have the eigenvalues $(\lambda^{t}_1,\lambda^{t}_2,\lambda^{t}_3)=(-1.85, -3, -4.7)$. For the matrix $\mathbf{D}$, we have the eigenvalues $(\lambda^{s}_1,\lambda^{s}_2,\lambda^{s}_3)=(0, -3, -5)$. As $\lambda_{min}=1.85<|\lambda^s_2|$, the synchronous pattern thus is judged as structurally unstable, i.e., it can not be generated by varying the coupling strength [Fig. 2(b)]. According to the theoretical predications, i.e., Eqs. (12) and (13), we have $\eta_1=(\sigma_c/\varepsilon)-\lambda_{min}=6.4$ and $\eta_2=\sigma_c/\varepsilon=8.3$, which agree with the numerical results well [Fig. 3(f)]. (The synchronous pattern associated to $\mathbf{S}_2$ also can be controlled, with the results similar to those of $\mathbf{S}_3$. Here we omit the details to save the space.)

\subsection{Power-grid network}

While the symmetries of simple networks can be discerned by inspection, the symmetries of large-size complex networks can only be identified with the help of some sophisticated techniques, e.g., by the tools of computational group theory \cite{CGT}. Besides, due to the existence of a large number of symmetries in complex networks, the scenario of cluster synchronization is much more complicated than the simple networks. For instance, the network may stay on a surprising state where one or more clusters lose synchronization while the remaining clusters are still synchronized, namely the phenomenon of isolated desynchronization \cite{CS:Pecora2014}.

To verify the feasibility of the proposed control method further, we next employ the Nepal power-grid network as the model \cite{Powergrid}, and investigate the controllability of its synchronous pattern. The structure of the Nepal power-grid is presented in Fig. 4(a), which consists of $15$ nodes (power stations) and $62$ links (power lines). For the sake of simplicity, we treat the network links as non-weighted and non-directed, i.e., $w_{ij}=w_{ji}=1$. By the technique of computational group theory, we are able to figure out all the network permutation symmetries (totally $86400$), and, according to the permutation orbits, partition the nodes into $5$ clusters: $V_1=\{1,2,3,4,5\}$, $V_2=\{6,7,8\}$, $V_3=\{9,10,11,12,13\}$, $V_4=\{14\}$, and $V_5=\{15\}$ \cite{CS:Pecora2014}. Among them, the $4$th and $5$th clusters are trivial, as they contains only a single node.

\begin{figure*}[tbp]
\begin{center}
\includegraphics[width=0.75\textwidth]{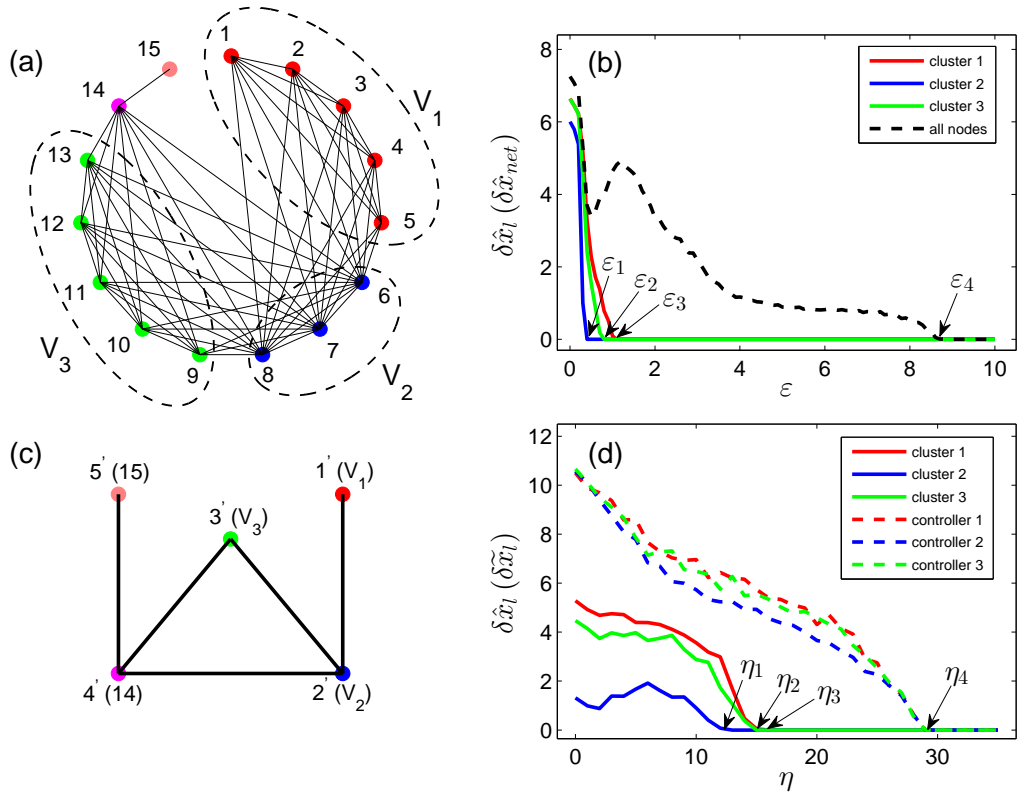}
\caption{(Color online) Controlling synchronous pattern in the Nepal power-grid network. The nodal dynamics and coupling function are the same to Fig. 3. (a) The network structure and the distribution of the clusters. The nodes are partitioned into three non-trivial clusters [$V_1=\{1,2,3,4,5\}$ (red), $V_2=\{6,7,8\}$ (blue), and $V_3=\{9,10,11,12,13\}$ (green)] and two trivial clusters [$V_4=\{14\}$ (pink) and $V_5=\{15\}$ (yellow)]. (b) The variations of the cluster synchronization errors, $\delta \hat{x}_{l}$ with $l=1,2,3$, and the network synchronization error, $\delta \hat{x}_{net}$, as a function of the coupling strength, $\varepsilon$. Clusters $2$, $3$ and $1$ are synchronized at $\varepsilon_1\approx 0.4$, $\varepsilon_2\approx 0.8$ and $\varepsilon_3\approx 1.1$, respectively. For $\varepsilon>\varepsilon_4\approx 8.9$, the network is globally synchronized. (c) The control network. Controller $l'$ is coupled unidirectionally to all oscillators in cluster $l$ of the original network. (d) For $\varepsilon=0.3$, the variations of the cluster synchronization errors, $\delta \hat{x}_l$, as a function of the controlling strength, $\eta$. Synchronization is induced in clusters $2$, $3$ and $1$ at $\eta_1\approx 13$, $\eta_2\approx 15$ and $\eta_3\approx 16$, respectively. Dashed lines: the variation of the cluster-controller synchronization errors, $\delta \hat{x}_l$, with respect to $\eta$. For $\eta>\eta_4\approx 29$, $\delta \tilde{x}_l=0$, indicating that the synchronous pattern is controlled to the state of the control network.
} \label{Fig4}
\end{center}
\end{figure*}

Employing still the Lorenz oscillator as the nodal dynamics, we plot in Fig. 4(b) the variation of the synchronization relationship among the oscillators as a function of the coupling strength, $\varepsilon$. Here, to better present the formation of the synchronous clusters, we monitor only the averaged synchronization error of the non-trivial clusters, $\delta \hat{x}_l=\sum_{i}\langle|x_i-\bar{x}_l|\rangle/n_l$, with $\bar{x}_l=\sum_i x_i/n_l$ the averaged state of the oscillators in cluster $l$. Clearly, if the $l$th cluster is synchronized, we have $\delta\hat{x}_l=0$. Fig. 4(b) shows that, with the increase of $\varepsilon$, cluster $1$ is firstly synchronized ($\varepsilon_1\approx 0.4$), followed by cluster $2$ ($\varepsilon_2\approx 0.8$) and then cluster $3$ ($\varepsilon_3\approx 1.1$). To exclude the possibility of global synchronization, we plot in Fig. 4(b) also the variation of the network-averaged synchronization error $\delta \hat{x}_{net}=\sum_{i}\langle|x_i-\bar{x}|\rangle/N$, with $\bar{x}=\sum_i x_i/N$ the network-averaged state. As $\delta \hat{x}_{net}=0$ at $\varepsilon_4\approx 8.9$, we thus confirm that cluster synchronization is generated in the range $\varepsilon\in (\varepsilon_3,\varepsilon_4)$. (Here cluster synchronization refers specifically to the state that all the non-trivial clusters are synchronized. If only part of the clusters are synchronization, we call it the state of isolated desynchronization [21]. The formation of this state will be discussed later in this section.)

The reference state to be controlled is generated by $\varepsilon=0.3<\varepsilon_1$, with which no synchronous cluster is formed on the network. Based on the cluster information, we can construct the control network shown in Fig. 4(c), which consists of $5$ controllers and $5$ weighted links. The coupling matrix of the control network reads [constructed according to Eq. (2)]
\begin{equation}
\mathbf{C}=\left(
  \begin{array}{ccccc}
      -3&   3  &   0&   0  &  0\\
       5&  -11 &   5&   1  &  0\\
       0&   3  &  -4&   1  &  0\\
       0&   3  &   5&  -9  &  1\\
       0&   0  &   0&   1  & -1\\
  \end{array}
\right).
\end{equation}
Controlling the reference state by the scheme proposed in Sec. II, we plot in Fig. 4(d) the variation of the averaged synchronization errors of the non-trivial clusters, $\delta \hat{x}_l$, as a function of the controlling strength, $\eta$. It is seen that as $\eta$ increases, the clusters $2$, $3$ and $1$ are synchronized at $\eta_1\approx 13$, $\eta_2\approx 15$ and $\eta_3\approx 16$, respectively. In particular, in the regime $\eta>\eta_3$, the synchronous pattern, which is unstable without the control, is successfully induced on the network. Increasing $\eta$ further to $\eta_c\approx 29$, in Fig. 4(d) it is seen that the synchronous pattern is controlled, i.e., each synchronous cluster is synchronized to its controller. (The control efficiency is measured by the cluster-controller synchronization errors, $\delta \tilde{x}_l=\sum_i \langle|x_i-x_{l'}|\rangle/n_l$, with $x_{l'}$ the state of the controller $l'$, and $i=1,\ldots,n_l$ the oscillators in cluster $l$.)

The critical controlling strengths, $\eta_3$ and $\eta_c$, can also be analyzed theoretically. According to the network cluster information, we have the following permutation matrix ($r_{ij}=r_{ji}=1$ if nodes $i$ and $j$ belong to the same cluster, and $r_{ij}=0$ otherwise)
\begin{equation}
\mathbf{R}=\left(
  \begin{array}{ccccccccccccccc}
      0&   1&   1&  1&  1&   0&   0&  0&   0&  0&  0&  0&  0&   0&  0\\
      1&   0&   1&  1&  1&   0&   0&  0&   0&  0&  0&  0&   0&  0&  0\\
      1&   1&   0&  1&  1&   0&   0&  0&   0&  0&  0&  0&   0&  0&  0\\
      1&   1&   1&  0&  1&   0&   0&  0&   0&  0&  0&  0&   0&  0&  0\\
      1&   1&   1&  1&  0&   0&   0&  0&   0&  0&  0&  0&   0&  0&  0\\
      0&   0&   0&  0&  0&   0&   1&  1&   0&  0&  0&  0&   0&  0&  0\\
      0&   0&   0&  0&  0&   1&   0&  1&   0&  0&  0&  0&   0&  0&  0\\
      0&   0&   0&  0&  0&   1&   1&  0&   0&  0&  0&  0&   0&  0&  0\\
      0&   0&   0&  0&  0&   0&   0&  0&   0&  1&  1&  1&   1&  0&  0\\
      0&   0&   0&  0&  0&   0&   0&  0&   1&  0&  1&  1&   1&  0&  0\\
      0&   0&   0&  0&  0&   0&   0&  0&   1&  1&  0&  1&   1&  0&  0\\
      0&   0&   0&  0&  0&   0&   0&  0&   1&  1&  1&  0&   1&  0&  0\\
      0&   0&   0&  0&  0&   0&   0&  0&   1&  1&  1&  1&   0&  0&  0\\
      0&   0&   0&  0&  0&   0&   0&  0&   0&  0&  0&  0&   0&  1&  0\\
      0&   0&   0&  0&  0&   0&   0&  0&   0&  0&  0&  0&   0&  0&  1\\
  \end{array}
\right),
\end{equation}
based on which we can obtain the transformation matrix $\mathbf{T}$. By $\mathbf{T}$, we then are able to transform the coupling matrix into the blocked form, with
\begin{equation}
\mathbf{B}=\left(
  \begin{array}{cccccccccc}
     -8&   0&   0  &    0 & 0&   0&   0  &    0 &   0  &    0\\
      0&   -8&   0  &    0 & 0&   0&   0  &    0 &   0  &    0\\
      0&   0&   -8  &    0 & 0&   0&   0  &    0 &   0  &    0\\
      0&   0&   0  &    -8 & 0&   0&   0  &    0 &   0  &    0\\
      0&   0&   0  &    0 & -14&   0&   0  &    0 &   0  &    0\\
      0&   0&   0  &    0 & 0&   -14&   0  &    0 &   0  &    0\\
      0&   0&   0  &    0 & 0&   0&   -9  &    0 &   0  &    0\\
      0&   0&   0  &    0 & 0&   0&   0  &    -9 &   0  &    0\\
      0&   0&   0  &    0 & 0&   0&   0  &    0 &   -9  &    0\\
      0&   0&   0  &    0 & 0&   0&   0  &    0 &   0  &    -9\\
  \end{array}
\right),
\end{equation}
and
\begin{equation}
\mathbf{D}=\left(
  \begin{array}{ccccc}
    -9&   1&   1.73&   0&   2.24\\
    1&   -1&        0&   0&   0\\
    1.73 &  0  & -11&   3.87&   3.87\\
    0&   0&   3.87&   -3&   0\\
    2.24    &  0  & 3.87&   0&   -4\\
  \end{array}
\right).
\end{equation}
The smallest eigenvalue of $\mathbf{B}$ is $\lambda_{min}=8$, which, according to Eq. (12), gives the critical controlling strength for inducing the synchronous pattern, $\eta_3\approx 20$. Meanwhile, according to Eq. (13), we also have the critical controlling strength for controlling the synchronous pattern to the state of the control network, $\eta_4\approx 28$. As depicted in Fig. 4(d), these theoretical results are  agreement well with the numerical results.

Besides the critical strengths $\eta_3$ and $\eta_4$, the other two critical controlling strengths observed in numerical simulations, i.e., $\eta_1$ and $\eta_2$, can also be analyzed. Noticing that $\mathbf{B}$ can be rewritten in the blocked form
\begin{equation}
\mathbf{B}=\left(
  \begin{array}{ccc}
    \mathbf{B}_1     &  0     &     0   \\
   0     &   \mathbf{B}_2     &      0   \\
0     &  0     &           \mathbf{B}_3   \\

  \end{array}
\right),
\end{equation}
with $\mathbf{B}_l$ the transverse space of the $l$th (non-trivial) cluster. [Please note that for the general complex networks, the matrix $\mathbf{B}$ has the blocked form shown in Eq. (28). But for the specific network of Nepal power-grid, $\mathbf{B}$ is diagonal.) According to the scenario of cluster synchronization depicted in Fig. 2, the $l$th cluster is synchronized when $\sigma_{l,1}=\varepsilon|\lambda^t_{l,1}|>\sigma_c$, with $\lambda_{l,1}$ the largest eigenvalue of $\mathbf{B}_l$. When control is added, $\lambda^t_{l,1}$ is replaced by $\lambda^t_{l,1}-\eta$ (as analyzed in Sec. III). The synchronization condition thus becomes $\varepsilon|\lambda^t_{l,1}-\eta|>\sigma_c$. That is, to make the $l$th cluster synchronizable (despite the synchronization relations of the remaining oscillators), the controlling strength should be larger to
\begin{equation}
\eta_l=(\sigma_c/\varepsilon-|\lambda^t_{l,1}|).
\end{equation}
For the matrices $\mathbf{B}_2$ and $\mathbf{B}_3$ in Eq. (28), we have $\lambda^t_{1,1}=-14$ and $\lambda^t_{2,1}=-9$, respectively. According to Eq. (29), we thus have $\eta_1=14$ (cluster $1$ is synchronized) and $\eta_2=19$ (cluster $2$ is synchronized), which agree with the numerical results very well (numerically we have $\eta_1\approx 13,\eta_2\approx 15$).

\subsection{Experimental study}

Can synchronous patterns be controlled in realistic systems? In our theoretical and numerical studies, it is assumed that the nodal dynamics is identical and the network structure is of perfect symmetry. In a realistic situation, parameter mismatch and noise perturbations are unavoidable. To check the feasibility of the control method to realistic networks, we finally study experimentally the control of cluster synchronization in network of coupled electronic circuits. Specifically, we adopt still the network structure of Fig. 3(a), but employ the Hindmarsh-Rose (HR) neuronal circuit as the nodal dynamics \cite{HR:model,HR:circuit}. The HR circuit is described by equations
\begin{equation}
 \begin{cases}
\dot{x} & =y-ax^3+bx^2-z+I_{e},\\
\dot{y} & =c-dx^2-y,\\
\dot{z} & = r[s(x+1.6)-z],
 \end{cases}
\end{equation}
with $(a,b,c,d,r,s)$ the system parameters, and $I_e$ the external forcing current. This model has been widely employed in literature for modeling the firing activities of neurons, and its experimental realizations by electronic circuits have been well designed and investigated. Here, we adopt the circuit diagram designed in Ref. \cite{HR:circuit}, and couple the circuits through the $x$ variable, i.e., $\mathbf{H}([x,y,z]^T)=[x,0,0]^T$. We set the system parameters as $(a,b,c,d,r,s)=(1,3,1,5,6\times 10^{-3},4)$, and choose the external current $I_e=320\mu A$, with which the circuit shows the chaotic motion \cite{HR:model}. The whole experimental process is controlled and monitored by the virtual interface of the software MULTISIM 12.0.

\begin{figure}[tbp]
\begin{center}
\includegraphics[width=0.8\linewidth]{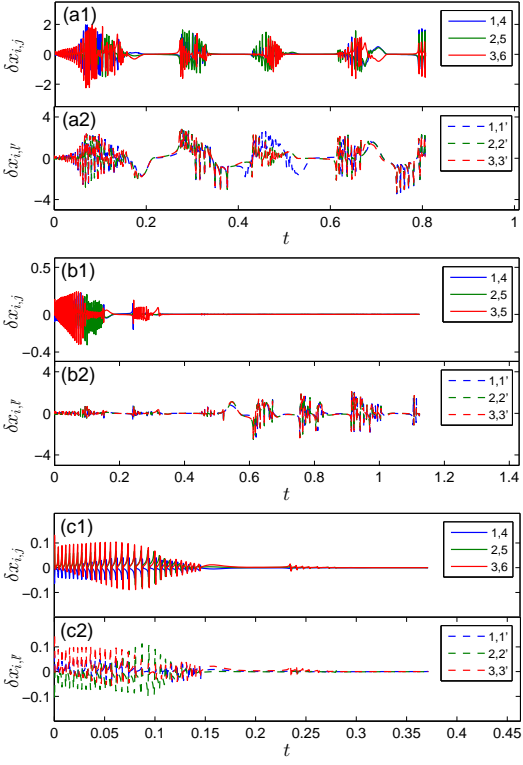}
\caption{(Color online) Experimental studies of controlling synchronous pattern in network of coupled chaotic HR circuits. The network structure is the same to Fig. 3(a), and the coupling strength is fixed as $\varepsilon=1.53$. The control network is the same to Fig. 3(e). The variations of the synchronization relationship among the oscillators (controllers) for different controlling strengths. (a) $\eta=1.0$. No synchronous cluster is formed. (b) $\eta=4.5$. Cluster synchronization is induced but is not controlled. (c) $\eta=6.9$. Cluster synchronization is both induced and controlled. $\delta \mathbf{x}_{i,j}=\mathbf{x}_i-\mathbf{x}_j$ is the synchronization error between circuits $i$ and $j$. $\delta \mathbf{x}_{i,l'}=\mathbf{x}_i-\mathbf{x}_l'$ is the control error between circuit $i$ and controller $l'$.} \label{Fig_4}
\end{center}
\end{figure}

To demonstrate, we choose Fig. 3(e) as the control network. That is, we are trying to induce and control the network to the synchronous pattern associated to $\mathbf{S}_3$, which, according to the eigenvalue analysis, is structurally unstable [Fig. 2(b)]. We fix the coupling strength among the circuits in the original network as $\varepsilon=1.53$, with which no synchronization is observed between any two circuits. Using this non-synchronous state as the reference state, we activate the control and record the synchronization relationship among the circuits under different controlling strengths, $\eta$. The typical results are plotted in Fig. 5. When $\eta$ is too weak, e.g., $\eta=1.0$ in Fig. 5(a), it is seen that the circuits in the original network are neither synchronized in pairs [Fig. 5(a1)] nor controlled to their corresponding controllers [Fig. 5(a2)]. Increasing the controlling strength to $\eta=4.5$ [Fig. 5(b)], it is seen that circuits $1$, $2$ and $3$ are synchronized with circuits $4$, $5$ and $6$, respectively, i.e., the cluster-synchronization state is induced [Fig. 5(b1)]. However, as depicted in Fig. 5(b2), the synchronized pairs are not controlled to their corresponding controllers, i.e., the cluster-synchronization state is not controlled. Increasing $\eta$ further to $6.9$ [Fig. 5(c)], it is seen that not only the cluster-synchronization state is induced in the original network [Fig. 5(c1)], but also it is controlled to the control network [Fig. 5(c2)].

For the HR circuits described by Eq. (30), in the experimental simulation, we find that the largest Lyapunov exponent is negative in the regime $\sigma>\sigma_c\approx 9.5$. As the structures of the original and control networks are the same to Figs. 3(a) and (e), the eigenvalues of the transverse and synchronous modes therefore are also given by Eqs. (22) and (23), respectively. According to the theoretical predications, i.e., Eqs. (12) and (13), we thus have $\eta_1\approx 4.4$ (for inducing cluster synchronization) and $\eta_2\approx 6.2$ (for controlling cluster synchronization). As depicted in Fig. 5, these theoretical predictions fit the experimental results very well.

\section{DISCUSSIONS AND CONCLUSION}

We would like to make the following remarks. Firstly, the current study is inspired by the recent progress of cluster synchronization in complex networks \cite{CS:BAO,CS:CSZ,CS:OTT2007,CS:Dahms,CS:Nicosia,PATTERNCONTROL:WXG,CS:WXG2014,CS:Pecora2014}. In particular, in Ref. \cite{CS:Pecora2014} the authors have proposed a numerical method for identifying the topological clusters, which paves the way to investigating cluster synchronization in large-size complex networks. However, different from previous studies which emphasizes the emergence of cluster synchronization in autonomous systems, here we focus on the control of cluster synchronization by an elaborately designed small-size network. As we have shown, with the help of the control network, not only the dynamically unstable patterns can be generated on the network (which are rare to be observed for complex networks), but also the structurally unstable ones. Regarding the important implications of synchronous patterns to the functioning of many realistic complex networks, the control method proposed in the present work may have broad applications.

Secondly, the present work is essentially different from the existing studies of controlling network synchronization. In parallel with the investigations of network synchronization, in the past years considerable attentions have been also given to the synchronization of complex networks driven by an externally added controller, namely the scheme of pinning synchronization \cite{PIN-1,PIN-2,PIN-3}. In the general picture of pinning synchronization, a controller, which has the same dynamics as the isolated network node, is coupled to some of the network nodes unidirectionally (pinning coupling), and the task is to control the network to the \emph{uniform} state of global synchronization \cite{PIN-1,PIN-2,PIN-3}. As the controller has the same dynamics and coupling function as the network nodes, the study of pinning synchronization is essentially a problem of global synchronization by treating the controller as an additional node to the exiting network. Different from pinning synchronization, in our present work the control network itself is a spatially extended system, which is designed according to the network symmetries. To control the network to different synchronous patterns, different control networks should be designed. More importantly, here the targeting state is spatially \emph{non-uniform}, i.e., the synchronous pattern, instead of the uniform state of global synchronization. For this difference, the stability analysis of cluster synchronization is essentially different from that of global synchronization, as shown in Sec. II.

Thirdly, the proposed method can be also used to control the chimera-like synchronization states \cite{CHIMERA-1,CHIMERA-2}. This interesting state can be observed in Fig. 4(b) in the range $\varepsilon\in(\varepsilon_1,\varepsilon_2)$, where only the nodes within the $1$st cluster are synchronized, while the remaining nodes of the network are desynchronized. As such, the synchronous and non-synchronous motions are coexisting in the network. This feature is very similar to that of the chimera state observed recently in the lattices of coupled periodic oscillators \cite{CHIMERA-1,CHIMERA-2}, and is termed as isolated desynchronization in Ref. \cite{CS:Pecora2014}. As depicted in Fig. 2, the isolated-desynchronization state could be unstable, due to either the weak coupling strength [Fig. 2(a)] or the network structure [Fig. 2(b)]. To control the isolated-desynchronization states, we can simply replace the control network by a single controller, and couple it unidirectionally to all nodes within the targeting cluster. The stability analysis will be identical to that of cluster synchronization given in Sec. II, except that the synchronous manifold is now defined by the controller, and only the transverse modes of the targeting cluster is concerned.

Finally, the proposed control method is applicable to the general complex networks. As the underlying mechanism of pinning control is governed by synchronization, the proposed control method, in principle, can be applied to any network showing synchronization behaviors. The generality of the control method has been verified by additional simulations. In particular, we have successfully applied this method to control complex networks of larger sizes ($N=50$ and $100$) and other types of nodal dynamics (chaotic logistic map and R\"{o}ssler oscillators) (not shown). Besides, after some modifications, the control method could be also generalized to controlling cluster synchronization in complex networks consisting of non-identical oscillators. This idea is inspired by the isolated-desynchronization state described above \cite{CS:Pecora2014}, where a synchronous cluster is appeared on the desynchronization background. Given nodes within the same cluster have the same dynamics, the network symmetry will be still satisfied, and the same control method will be applies. We hope to concrete this idea by further studies.

In summary, we have proposed a general framework for controlling synchronous patterns in complex networks, including how to design the control network based on the information of network symmetries, the control mechanism, and the formula of the critical controlling strength. The efficiency of the control method is justified by numerical simulations on both artificial and realistic network models, and its feasibility is verified by realistic experiment of networked electric circuits. Our studies highlight the fact that, like other types of collective behaviors, synchronous patterns in complex networks are also controllable.

This work was supported by the National Natural Science Foundation of China under the Grant No.~11375109 and by the Fundamental Research Funds for the Central Universities under Grant No.~GK201303002.

\end{document}